\documentclass[12pt]{article}
\usepackage{amsmath,amssymb,color}
\usepackage{cite,epsf}

\usepackage{graphicx,array} 
\newcommand{\be}{\begin{equation}}
\newcommand{\ee}{\end{equation}}
\newcommand{\beq}{\begin{equation}}
\newcommand{\eeq}{\end{equation}}
\newcommand{\bea}{\begin{eqnarray}}
\newcommand{\eea}{\end{eqnarray}}

\newcommand{\ba}{\begin{eqnarray}}
\newcommand{\ea}{\end{eqnarray}}

\begin{document}

\begin{titlepage}
\vspace{10pt}
\hfill
{\large\bf HU-EP-09/47}
\vspace{20mm}
\begin{center}

{\Large\bf  Some comments on spacelike minimal surfaces\\[2mm] with null polygonal
  boundaries in $AdS_m$}

\vspace{45pt}

{\large Harald Dorn
{\footnote{dorn@physik.hu-berlin.de
 }}}
\\[15mm]
{\it\ Institut f\"ur Physik der
Humboldt-Universit\"at zu Berlin,}\\
{\it Newtonstra{\ss}e 15, D-12489 Berlin, Germany}\\[4mm]

\vspace{20pt}

\end{center}
\vspace{10pt}
\vspace{40pt}

\centerline{{\bf{Abstract}}}
\vspace*{5mm}
\noindent
We discuss some geometrical issues related to spacelike minimal surfaces
in $AdS_m$ with null polygonal boundaries at conformal infinity. In particular
for $AdS_4$, two holomorphic input functions for the Pohlmeyer
reduced system are identified. This system contains two coupled differential
equations for 
two functions $\alpha (z,\bar z)$ and $\beta (z,\bar z)$, 
related to curvature and torsion of the surface. Furthermore, we conjecture
that, for a polynomial choice of the two holomorphic functions,
the relative positions of their zeros encode the conformal invariant data of
the 
boundary null $2n$-gon.
\vspace{15pt}
\noindent
\end{titlepage}
\newpage
\section{Introduction}
The $N$-point MHV gluon scattering amplitude at strong coupling in ${\cal N}=4$
super Yang-Mills is related to a string worldsheet in $AdS_5$, approaching a
$N$-sided polygon spanned by the lightlike momenta of the scattering process
on the conformal boundary of $AdS_5$ \cite{am}. To find the corresponding 
minimal surface, is a difficult Plateau-like problem, and not much is known
for the case of a generic null polygonal boundary. For the tetragon 
the surface has been constructed explicitely and, calculating the regularized
area, the conjecture has been checked for the case of the four-point amplitude
\cite{am}. There has been also an interesting discussion of the limit
of a large number of polygon sides, which led to the conclusion that the
BDS ansatz \cite{bds} for gluon amplitudes breaks down \cite{am2}. 

By inspection, this tetragon surface is spacelike and flat. In our
previous work \cite{djw} we have proven, that besides isometry
transformations, there are no other {\it flat}
spacelike minimal surfaces. The flatness was an issue, because there exist
flat timelike surfaces beyond the Wick rotated version of the tetragon surface,
and because flatness would have simplified the explicit construction of the
wanted surfaces.  

Besides numerical work \cite{dobashi} and a screening of some
solutions of the relevant differential equations with respect to
their boundary behaviour \cite{ss}, the only systematic progress
has been made in \cite{am3}. Seen geometrically, the crucial 
new insight of this work is the fact, that in $AdS_3$ null N-gons ($N>4$)
only can arise as boundaries, if the second fundamental form
has zeros on the surface. In more detail, it has been shown, that the conformal
invariant data of the wanted boundary null polygon
are  in one to one correspondence to the relative position of zeros
of a holomorphic polynomial parameterizing the second fundamental form.
This includes a certain boundary condition for the scale factor of the induced
metric in conformal coordinates. Although the surfaces could not be
constructed explicitely, the authors of \cite{am3} where able to
calculate the regularized area of the minimal surface related to an octagon
in 2-dimensional Minkowski space. In this way an explicit formula for
the remainder function, which describes the part not fixed by anomalous
dual conformal Ward identities, has been found for the 8-point amplitude.

Taking place in a $\mathbb{R}^{1,1}$ subspace of physical 4-dimensional
Minkowski space is of course a degenerated case for a $N$ particle
scattering. The aim of the present paper is to add some observations
which could be helpful to extend the strategy of \cite{am3} to less
degenerated or even the generic kinematics. 

The paper is organized as follows. To fix the notation, we summarize in
section 2 
some necessary formulae from \cite{djw}. Furthermore,
this section contains a more elaborate discussion of the scalar invariants
characterizing minimal surfaces in $AdS_m$ and a counting of the number of
independent cross ratios formed out of the position of the vertices
of the null N-gons in arbitrary dimensional Minkowski space. Section 3
describes in some detail geometrical issues in $AdS_3$, related to the 
identification of the one holomorphic polynomial carrying all the information
about the boundary. In section 4 we turn to $AdS_4$ and identify two
holomorphic functions which serve in the Pohlmeyer reduced system
as an input for a coupled system of two differential equations controlling
curvature and torsion of the minimal surface. The structural similarity
to the $AdS_3$ case and the matching of the numbers of parameters allows us
to formulate in section 5 the conjecture, that now the two holomorphic
functions of section 4 carry all information about the boundary. In section 6
we make some remarks on the more complicated full problem in $AdS_5$.    
\section{The general framework for minimal surfaces\\ in $AdS_m$}      
Realizing $AdS_m$ (with coordinates $X^k$) as a hyperboloid in
$\mathbb{R}^{2,m-1}$ (coordinates $Y^N$)
and choosing conformal coordinates on the surface, one gets as the 
minimal surface condition
\beq
\partial \bar{\partial}Y^N(X(z))~-~\partial Y^K\bar{\partial}Y_K \ Y^N~=~0~.
\label{eom}
\eeq 
The choice of conformal coordinates gives the additional condition
\beq
\partial Y^N\partial Y_N~=~\bar{\partial}Y^N\bar{\partial}Y_N~=~0~,
\eeq
where $\partial ,~\bar{\partial}$ are defined by $\partial =\partial
_{\sigma}+\partial_{\tau},~~\bar{\partial} =\partial 
_{\sigma}-\partial_{\tau}$ for timelike surfaces and by $\partial =\partial
_{\sigma}-i\partial_{\tau},~~ \bar{\partial} =\partial 
_{\sigma}+i\partial_{\tau}$ for spacelike surfaces.

One now extends the vectors $Y,\partial
Y,\bar{\partial}Y$ to a basis  of $\mathbb{R}^{2,m-1}$ \cite{dvs,jevicki}
\beq
\{e_N\}~=~\{Y,\partial
Y,\bar{\partial}Y,B_4,\dots ,B_{m+1}\}~.\label{basis}
\eeq  
The orthonormal vectors $B_a$ pointwise span the normal space of the surface
inside $AdS_m$.  
Due to the hyperboloid condition, $Y$ is timelike. For timelike surfaces a
further 
timelike  vector is parallel to the surface, hence the normal space has to be
positive definite. In contrast, for spacelike surfaces the second timelike
vector has to be in the normal space. We choose it to be $B_4$. With
($a,b=4,\dots ,m+1$) 
\beq
h_{ab}~=~\delta _{ab}~\mbox{or}~\eta
_{ab},~~~\mbox{for timelike or spacelike surface,}\label{h}
\eeq
we require
\beq 
(B_a,B_b)~=~h_{ab}~,~~(B_a,Y)~=~(B_a,\partial Y)~=~(B_a,\bar{\partial}Y)~=~0~.
\label{transversal}
\eeq
Moving the basis (\ref{basis}) along the surface one
gets
\beq
\partial \ e_N~=~A_N^{~~K}~e_K~,~~~~~\bar{\partial}\  e_N~=~\bar
A_N^{~~K}~e_K~. \label{evol}
\eeq
Introducing
\bea
\alpha(\sigma ,\tau)&=&\log (\partial Y,\bar{\partial} Y)\label{alpha}\\
u_a(\sigma ,\tau)&=&(B_a,\partial \partial Y)~,~~~\bar u_a (\sigma
,\tau)~=~(B_a,\bar{\partial}\bar{ \partial }Y)~,\nonumber\\
A_{ab}(\sigma,\tau)&=&(\partial B_a,B_b)~,~~~~~~~~~\bar A_{ab}(\sigma,\tau)~=~(\bar{\partial }B_a,B_b)~, 
\label{u}
\eea 
and using (\ref{eom}), (\ref{transversal}) one can give eqs. (\ref{evol}) a more detailed form
\bea
\partial Y&=&~~~~~~~~~~\partial Y\nonumber\\
\partial\partial Y&=&~~~~~~\partial\alpha\partial Y~~~~~~~~~~~~~~~+~u^bB_b\nonumber\\
\partial\bar{\partial}Y&=&e^{\alpha}Y\nonumber\\
\partial B_a&=&~~~~~~~~~~~~-e^{-\alpha}~u_a\bar{\partial}Y~+~A_a^{~~b}B_b~,\label{master}
\eea
as well as the equations which one gets by the replacements
$\partial\leftrightarrow\bar{\partial}$, $u_a\rightarrow \bar u_a$, $
 A_a^{~~b}\rightarrow \bar A_a^{~~b}$.
\footnote{Note that for timelike surfaces $u$ and $\bar u$ as well as $A$ and
$\bar A$ are real. On the other side, for spacelike surfaces $u$ and $A$ are
complex, and then the bar means complex conjugation.} Indices on $u,\bar u$ and
$A,\bar A$ are raised and lowered with the normal space metric $h$, see
eq. (\ref{h}). $A$ and $\bar A$ with both indices downstairs are antisymmetric.

Then, the integrability condition $\partial\bar{\partial}e_N=\bar{\partial}
\partial e_N$ for eq. (\ref{evol}) gives
\bea
\partial\bar{\partial}\alpha-e^{-\alpha}u^b\bar
u_b-e^{\alpha}&=&0~,\label{gauss}\\
\partial \bar u_a - ~A_a^{~~b}\bar u_b&=&0~,~~~~~~~~~~ \bar{\partial}u_a -
~\bar A_a^{~~b}u_b~=~0~,\label{codazzi}\\
e^{-\alpha}\left (\bar u_au^b-u_a\bar u^b\right )&=&F_a^{~b}~,\label{ricci}
\eea
with
\beq
F_a^{~b}~=~\partial \bar A_a^{~~b} -
\bar{\partial} A_a^{~~b}+\bar A_a^{~~c}  A_c^{~~b}- A_a^{~~c} \bar
A_c^{~~b}~.\label{F}
\eeq
$A$ appears as a gauge field with the related field strength $F$
($A\in so(1,m-3)$ for spacelike surfaces and $A\in so(m-2)$ for timelike
surfaces). 
 
Here, a comment on the geometrical meaning of our quantities $\alpha ,~u,~A$ is
in order. Since we are using conformal coordinates,
\beq
R~=~-2\ e^{-\alpha}       \ \partial\bar{\partial}\alpha\label{R}
\eeq
is the curvature scalar on our surface. $u,~\bar u$ parameterize the
second fundamental forms 
$l^{c}_{\mu\nu}=(B^c,\partial _{\mu}\partial_{\nu}Y)$ with built in minimal surface
condition  $l^{c~\mu}_{~~\mu}=0$.
 
The matrices $A,~\bar A$ in (\ref{codazzi}),(\ref{ricci})
describe the torsion of the surface (for $AdS_m,~m\geq\nolinebreak
4$). Eqs.(\ref{gauss})-(\ref{ricci}) are the Gau{\ss}, Codazzi-Mainardi and
Ricci equations specialized to minimal surfaces in conformal coordinates. In
the physics literature their described derivation is often called Pohlmeyer
reduction \cite{pohl}. \\

After this general discussion with spacelike and timelike in parallel, we now
restrict to spacelike minimal surfaces. For more comments on timelike
surfaces see e.g. \cite{djw,j}.

To form out of $F$ scalar invariants with respect to $SO(1,m-3)$
transformations in the 
normal space, one has at ones disposal the traces tr$(F^n)$.
Due to the special structure imposed for minimal surfaces by the Ricci
equation (\ref{ricci}), one finds
\beq
\mbox{tr}(F^{2n+1})~=~0~,~~~~~~~\mbox{tr}(F^{2n})~=~2^{1-n}~\Big
(\mbox{tr}(F^2)\Big )^n~,\label{ftrace}
\eeq 
which means that tr$(F^2)$ is the only independent $SO(1,m-3)$ invariant. 
To get a quantity invariant in addition with respect to conformal coordinate
changes on the surface $z\mapsto \zeta (z)$, one has to compensate the
transformation of tr$(F^2)$ by that of a suitable power of $e^{-\alpha}$.
Following \cite{djw} we introduce 
\beq
T~:=~\frac{1}{2}~e^{-2\alpha}~\mbox{tr}(F^2)~.\label{torsion}
\eeq 
One could form invariants also directly out of $u$ and $\bar u$, i.e.
\footnote{The conformal transformations of $uu$ or $\bar u\bar u$
alone cannot be compensated by a power of $e^{-\alpha}$.}
\beq
K~:=~e^{-2\alpha}~u_a\bar u^a~,~~~~L~:=~e^{-4\alpha}~u_au^a~\bar u _b\bar u^b~.
\label{KL}
\eeq
However, they contain no new information since due to the Ricci equation
(\ref{ricci}) 
\beq
T~=~K^2~-~L~,\label{TKL}
\eeq 
and by the Gau{\ss} equation (\ref{gauss})
\beq
R~+~2K~+~2~=~0~.\label{RK}
\eeq
This shows that for minimal surfaces, in generic $AdS_m$, the scalar curvature 
$R$ and the torsion invariant $T$ are the only independent scalar invariants
built  out of the surfaces induced metric, curvature tensor and the torsion $F_a^{~b}$. This
of course refers to invariants without derivatives. A further outcome of this
discussion is a universal inequality \cite{djw} arising from (\ref{TKL}),
(\ref{RK}) and the semi-definiteness of $L\geq 0$ 
\beq
\frac{(R+2)^2}{4}~-~T~ \geq 0~.\label{bound}
\eeq
Surfaces in $AdS_3$ have no torsion, i.e. then in the equations above one has
to set always $T=0$.  
For the discussion of the sign of $T$
 for $m>3$ we define $u_{\pm}=u_5\pm u_4$, $\vec
u=(u_6,u_7,\dots )$, $\vec u=\vec a +i\vec b$ and get
\beq
(u_k\bar u^k)^2-u_ku^k~\bar u_l\bar u^l~=~-(\mbox{Im}(u_+\bar
u_-))^2+4~\mbox{Im}(\vec u\ \bar u_+)\mbox{Im}(\vec u\ \bar u_-)+4~(\vec a^{2}\
 \vec b^{2}-(\vec a\ \vec b)^2)~.\label{signT}
\eeq 
The first term on the r.h.s. is always $\leq 0$, the last one always $\geq 0$.
For $AdS_4$, there is
no $\vec u=\vec a +i\vec b$. Hence only the first term is present and we get
$T\leq 0$. For $AdS_m,~m\geq 5$ both signs are possible, but due
(\ref{bound}) $T$ is nevertheless bounded from above. \footnote{Note that for
  timelike 
  surfaces one always has $T\leq 0$. If they are minimal 
  (\ref{bound}) is valid, too.}\\

The minimal surfaces needed in the Alday-Maldacena conjecture have to solve
a Plateau-like problem, i.e. they have to extend
to infinity and to approach a null polygonal at the conformal boundary of
$AdS_m$. Since isometries of $AdS_m$ act as conformal transformations on the
boundary, the relevant boundary data are encoded in the conformal invariants
of null polygonals in $(m-1)$-dimensional Minkowski space. These are given
by  cross ratios formed out of the positions of the cusps of the polygon.
Counting the number of coordinates, taking into account the constraints
set by the null condition for the sides of the polygon and subtracting the
number of parameters of the conformal group one gets for the
number of independent cross ratios for a null $N$-gon in $D=m-1$ dimensions
\beq
C~=~\mbox{max}\left \{0~,~N(D-1)~-~\frac{(D+1)(D+2)}{2}\right \}~.\label{crossratios}
\eeq
\section{Spacelike minimal surfaces in $AdS_3$}
In this most simplest case there is no chance for torsion, there is only one
complex valued function $u=u_4$, parameterizing the single second fundamental
form.  Equation (\ref{codazzi}) forces $u$ to be holomorphic and
eq.(\ref{gauss}) becomes
\beq
\partial\bar{\partial}\alpha +e^{-\alpha}u\bar u -e^{\alpha}=0~.\label{3gauss}
\eeq
$u(z)$ behaves under conformal coordinate transformations $z\mapsto \zeta (z)$ 
as
\beq
u(z)~\mapsto ~ (\zeta '(z))^{-2}u(z)~.\label{u-conf}
\eeq 
If $u(z)\neq 0$ in some open set of the surface, then there one can transform
$u$ to a constant, e.g.  $u=1$. Now the choice of the conformal
coordinates in this open set is up to translations fixed completely, and one has to solve
the sinh-Gordon equation
\beq
 \partial\bar{\partial}\alpha -2\sinh \alpha =0~.\label{sinh}
\eeq 
Inserting the trivial solution $\alpha (z,\bar z)=0$ into the linear problem
(\ref{master}), to reconstruct the embedding of the surface, one gets (due to
good luck) the well-known tetragon solution of the Plateau-like problem under
investigation \cite{jevicki,djw,am3}. This explicit solution is defined in the
whole $z$-plane, and the boundary of $AdS_3$ is reached for $\vert z\vert
\rightarrow \infty $. 

From now we {\it assume} the existence of such a globally defined system of
conformal coordinates, with the property that $\vert z\vert
\rightarrow \infty $ is mapped to the null N-gon on the boundary of $AdS_m$
also for other cases, i.e. both higher $N$ or/and higher $m$. \footnote{As
  a shorthand we will call them global conformal coordinates.} One to one
conformal maps of the complex plane are given by the M\"obius group. Since in
addition we insist on the correlation of infinite $z$ to the boundary N-gon,
only translations and dilatations remain as a freedom for the choice
of the global conformal coordinates. In contrast to the infinite dimensional
local freedom (\ref{u-conf}), the required global property thus fixes the
conformal coordinates up to the choice of the origin and up to multiplication
with a constant. We also require, that in this coordinates all functions
appearing in (\ref{gauss})-(\ref{ricci}) are free of singularities at finite
$z$. 

The construction of \cite{am3} shows, that such global conformal coordinates
exist in $AdS_3$ also for $N>4$. Based on this, our assumption for
higher dimensional $AdS_m$ is justified for null polygons in the neighbourhood
of those degenerated to a location in a two-dimensional Minkowski space.
For null polygonal configurations far from the degenerated ones, one should
expect the possibility of branched minimal surfaces. Branched minimal surfaces
come into the game also in the classical Plateau problem in $\mathbb{R}^m$ if
one goes beyond $m=3$ \cite{osser}.

To handle minimal surfaces with boundary null $N$-gons
with even $N>4$ in $AdS_3$, the authors of  \cite{am3} allow zeros of
$u(z)$.  They start with a polynomial ansatz for $u(z)$ in global conformal
coordinates and are able to relate the data parameterizing the relative
position of the zeros of this polynomial in a bijective manner 
to the cross ratios of the boundary N-gon. For $AdS_3$, i.e. $D=2$, formula
(\ref{crossratios}) gives for an $(N=2n)$-gon $2(n-3)$ independent (real)
cross ratios. Therefore, the polynomial $u(z)$ has to be of degree $(n-2)$.

In contrast to the $2n-6$ cross ratios for a $2n$-gon degenerated
to live in  a $\mathbb{R}^{1,1}\subset \mathbb{R}^{1,3}$,  a generic
$2n$-gon in 4-dimensional Minkowski space via (\ref{crossratios}) has
$6n-15$ independent cross ratios. For a partial lift of the degeneracy
via a generic embedding in a $\mathbb{R}^{1,2}$ one has to handle
$4n-10$ cross ratios \footnote{For the octagon, discussed for
  $\mathbb{R}^{1,1}$ in detail in \cite{am3}, the numbers are 2, 9 and 6.}. 
To make at least partial progress beyond \cite{am3},
a natural step is the discussion of  minimal surfaces in $AdS_4$.\\

We close this section by some geometrical comments, which give another
motivation for the polynomial ansatz for $u(z)$ in \cite{am3}.  In terms of the
scalar 
invariants, discussed in the previous section, $u=0$ (at finite $\alpha$) 
implies $K=0$ and via (\ref{RK}) $R=-2$. This value of $R$ coincides with that
for the surface $\mathbb{H}^2$.  If $u=0$ in some open set, then there
all geodesics of the surface would be also geodesics of the embedding $AdS_3$ 
and the surface called totally geodesic. If $u=0$ at an isolated point,
then there all geodesics of the surface passing this point have
zero curvature in the sense of $AdS_3$.

In addition, there is a nice descriptive argument for the necessity of
zeros of $u$ for $N=2n>4$. Let us map $AdS_3$ to half of ESU$_3$, i.e. a
cylinder in $\mathbb{R}^3$ and consider the maximal symmetric null 2n-gon on
its boundary. Furthermore, we consider the geodesics on the surface
connecting the middle points of the opposite sides of the polygons.
In the case of the tetragon one finds by explicit calculations, that
these lines are also geodesics in the sense of $AdS_3$. Their image
in the ESU$_3$ are just the straight
lines (in the sense of $\mathbb{R}^3$) connecting the middle points of the
opposite sides of the polygons and crossing each other on the axis of the
cylinder.  

Due to the symmetry of the problem, one expects these straight
lines (in the sense of $\mathbb{R}^3$) to be geodesic both in the sense of the
surface and $AdS_3$ for $2n>4$, too. At a point
of a minimal surface with $u\neq 0$ at most two lines, geodesic both in the
sense of 
the surface as well in the sense of the embedding $AdS_3$, can cross (see
appendix).
This shows that starting from $n=3$ the crossing point must be a
(multiple) zero of $u$. As shown by the analysis of \cite{am3}, for the 
generic unsymmetric configuration the multiple zero is dissolved into
separated single zeros.

\section{Spacelike minimal surfaces in $AdS_4$}
Here the new degree of freedom relative to $AdS_3$ allows minimal surfaces
with torsion. Vice versa we expect torsion necessary to get surfaces
for null polygons winding in full $\mathbb{R}^{1,2}$.

We have an Abelian gauge group related to the surfaces normal space,
$SO(1,1\vert\mathbb{R})$. Denoting 
\beq
\phi~:=~A_4^{~5}=A_{45}=-A_{54}=A_5^{~4}~,~~~~~u_{\pm}~=~u_5~\pm ~u_4
\label{phiupm} 
\eeq
the Codazzi equations in coordinates decouple
\beq
\partial \bar u_{\pm}~\mp \phi\bar u_{\pm}~=~0~,~~~~\bar \partial u_{\pm}~\mp
\bar \phi u_{\pm}~=~0~.\label{deccodazzi} 
\eeq    
The gauge field $A_a^{~b}$ is related to the complex derivative $\partial$.
Hence it is $\in so(1,1\vert \mathbb{C}) $, which implies $\phi\in
\mathbb{C}$.  
Gauge transformations are described by 
\bea
\phi~\mapsto~\phi~+~\partial \omega &,&~~~\bar \phi~\mapsto~\bar \phi~+~\bar
\partial \omega ~,~~~~\omega~\in~\mathbb{R}~,\nonumber\\
u_{\pm}~\mapsto ~e^{\pm \omega}u_{\pm}~&,&~~~\bar u_{\pm}~\mapsto ~e^{\pm
  \omega}\bar u_{\pm}~.\label{pmgaugetrafo}
\eea
Now we parameterize 
\beq
\phi~=~\partial \eta ~,~~~\bar\phi~=~\bar \partial \bar \eta
~,~~~\eta\in\mathbb{C}\label{gaugepar}.
\eeq
Gauge transformations then look like
\beq
\eta (z,\bar z)~\mapsto~\eta(z,\bar z)~+~\omega (z,\bar z)~,~~~~\omega\in
\mathbb{R}~. \label{etagauge}
\eeq
There remains a gauge parameterization freedom
\beq
\eta (z,\bar z)~\mapsto ~\eta (z,\bar z)~+~\overline{\xi(z)}~,\label{repar}
\eeq
with holomorphic $\xi (z)$.
The inversion of (\ref{gaugepar}) is then
\beq
\eta (z,\bar z)~=~\frac{1}{\partial \bar{\partial}}~\bar{\partial}\phi (z,\bar
z)~+~\overline{\xi (z)}~.\label{inversion}
\eeq

Introducing the gauge invariants
\beq
\bar v_{\pm}~:=~e^{\mp \eta}~\bar u_{\pm}~,~~~v_{\pm}~:=~e^{\mp
  \bar\eta}u_{\pm}\label{vpm}
\eeq
and
\beq
\beta ~:=~-i~(\eta -\bar\eta)~,\label{eta}
\eeq
the equations (\ref{gauss})-(\ref{ricci}) take the form
\bea
\partial\bar{\partial}\alpha-\frac{e^{-\alpha}}{2}\big (e^{i\beta}v_-\bar
v_+~+~e^{-i\beta}v_+\bar v_-\big )
-e^{\alpha}&=&0~,\label{4gauss}\\
\partial \bar v_{\pm}&=&0~,~~~~~~~
\bar{\partial}v_{\pm}~=~0~,\label{4codazzi}\\ 
\partial\bar\partial\beta~+~\frac{e^{-\alpha}}{2i}\big (e^{i\beta}v_-\bar
v_+~-~e^{-i\beta}v_+\bar v_-\big )&=&0~.\label{4ricci}
\eea
Due to (\ref{4codazzi}), the two functions  $v_{\pm}$ have to be
holomorphic. They appear as an 
input in the two coupled equations for the real functions $\alpha$ and
$\beta$. This is similar to the situation in $AdS_3$, where one holomorphic
equation appears as an input into one equation for $\alpha$.

As an aside let us mention, that with $\gamma :=\alpha -i\beta$ the two
equations (\ref{4gauss}) and (\ref{4ricci}) can be combined into one
equation for a complex valued function $\gamma$
\beq
\partial\bar\partial\gamma~-~e^{-\gamma}v_-\bar v_+~-~e^{\frac{1}{2}(\gamma
  +\bar\gamma)}~=~0~.\label{gammaeq}
\eeq 

$\beta $ and $v_{\pm}$ are gauge invariant, but we have traded another
unphysical degree of freedom, the gauge parameterization freedom
(\ref{repar}). The two holomorphic functions $v_{\pm}$ transform
under (\ref{repar}) as
\beq
v_{\pm}(z)~\mapsto ~e^{\mp \xi (z)}~v_{\pm}(z)~,\label{v-repar}
\eeq 
and under conformal coordinate transformations $z\mapsto \zeta (z)$ as
\beq
v_{\pm}(z)~\mapsto ~ (\zeta '(z))^{-2}v_{\pm}(z)~.\label{v-conf}
\eeq 
We see that, although $v_{\pm}(z)$ transform, the position of their zeros has
an invariant meaning. 

To explore the consequences for the scalar invariant quantities, we
use (\ref{phiupm}), 
(\ref{vpm}), (\ref{KL}) and (\ref{TKL}) to get
\beq
L=e^{-4\alpha}~\vert v_-\vert ^2\vert v_+\vert ^2~,~~K=e^{-2\alpha}~\mbox{Re}\left (e^{i\beta}v_-\bar v_+ \right)~,
~~T=-e^{-4\alpha}~\left (\mbox{Im}\left (e^{i\beta}v_-\bar v_+ \right
      )\right )^2~.\label{4KT}
\eeq 
Therefore, zeros of $v_-$ or $v_+$ are zeros of $L,~K$ and $T$. However, zeros
of 
$T$ or of 
$K$ (via (\ref{RK}) points with $R=-2$) appear also at other points,
generically on a net of lines in the 
$z$-plane. Only for $L$  we have $L=0~\Leftrightarrow ~v_-=0$ or $v_+=0$. 
Note that at just these points the universal inequality (\ref{bound}) for
$R$ and $T$ is saturated.

In this section our main result is twofold. At first we showed, that the
function $\bar u_a\bar u^a$, which is holomorphic for all $AdS_m,~m\geq 3$,
factorizes  in the two independent holomorphic functions $v_+$ and $v_-$ in
the case of 
$AdS_4$. At second we formulated the equations for the Pohlmeyer reduced
system for $AdS_4$ with these two holomorphic functions as input data.
Based on this we will motivate in the next section a conjecture on the
construction of minimal surfaces with null polygonal boundaries.
\section{Null polygonal boundaries in the $AdS_4$ case}
To start with, we consider the situation where $v_{\pm}$ in some open subset
have a finite number of zeros. The fact that
they transform the same way under local conformal transformations
(\ref{v-conf}), but in an inverse way under the change of gauge
parameterization (\ref{v-repar}), could be used to bring them both into a
polynomial form. However, having chosen global conformal coordinates as
defined in the previous section, the transformation (\ref{v-conf}) is no
longer available. But nevertheless, this observation supports somehow the
expectation that, similar to the
$AdS_3$ case \cite{am3}, the wanted null polygonal boundaries are realized,
if one starts with two polynomials in $z$ as an input. 

Let $v_-(z)$ and $v_+(z)$ be two polynomials of degree $(n-2)$, whose
coefficients in front of the highest power is one. We take this as input in
the coupled system of differential equations for $\alpha (z,\bar z)$ and
$\beta (z,\bar z)$, i.e. eqs. (\ref{4gauss}) and (\ref{4ricci}). With the
boundary condition specified below we expect, that the solution for
$\alpha (z,\bar z),~\beta (z,\bar z)$, after solving the linear problem
(\ref{master}), generates a minimal surface with a null $2n$-gonal boundary
at $\vert z\vert\rightarrow\infty $. 

There is strong support for this guess from counting parameters. The relative
position of zeros of $v_{\pm}(z)$ is characterized by $2(n-2)-1$ complex
parameters, i.e.
$4n-10$ real parameters. This just matches the number of independent cross
ratios for a null $2n$-gon in $\mathbb{R}^{1,2}$, as identified in section 3.

The degenerated case $v_-=v_+=:v(z)$ leads to 
$~\partial\bar{\partial}\alpha-\vert v\vert^2~e^{-\alpha}\cos
\beta-e^{\alpha}=\nolinebreak 0$,\\$\partial\bar{\partial}\beta +\vert v\vert
^2~e^{-\alpha}\sin\beta = 0$. The second equation is then solved by $\beta
=\pi $, which puts the first equation in the form of the $\alpha$-equation
in $AdS_3$.

It remains to discuss the boundary condition for $\alpha $ and $\beta $. We
closely follow the line of reasoning used in \cite{am3} for $AdS_3$. There the
boundary condition for $\alpha $ is naturally found in the $w$-plane, related
to the $z$-plane via $dw=\sqrt {u(z)}~dz$. The $w$-plane is no global conformal
coordinate system, to cover the whole surface, one has to go in a Riemann
surface over the $w$-plane. However, one has $\vert
w\vert\rightarrow\infty\Leftrightarrow \vert z\vert\rightarrow\infty$. The
transformed entries in the Gau{\ss} equation 
are $\hat u(w)=1$ and $\hat {\alpha}(w,\bar w)=\alpha (z,\bar z)-\log \vert
u(z)\vert~.$   
Reasoning, that for $\vert w\vert\rightarrow\infty $ the surface should behave
as the known tetragon solution, one gets the boundary condition $\hat
\alpha\rightarrow 0$. Translated back into the $z$-plane this means
$\alpha (z,\bar z)=\log \vert u(z)\vert +o(1)$ at $\vert z\vert
\rightarrow\infty $. 

For $AdS_4$ we define
\beq
dw~=~(v_+v_-)^{\frac{1}{4}}~dz~.\label{wplane}
\eeq  
Then the transformed entries for (\ref{4gauss}) and (\ref{4ricci}) are
\bea
\hat v_+(w)&=&\left (\frac{v_+(z)}{v_-(z)}\right )^{\frac{1}{2}}~,~~~~~~\hat 
v_-(w)~=~\left (\frac{v_-(z)}{v_+(z)}\right )^{\frac{1}{2}}~,\nonumber\\
\hat{\alpha}(w,\bar w)&=&\alpha (z,\bar z)~-\frac{1}{2}\log \vert v_+(z)v_-(z)\vert
~,~~~~~~
\hat {\beta}(w,\bar w)~=~\beta (z,\bar z)~.\label{wz}
\eea
Contrary to the $AdS_3$ case, in the generic situation, the transformed $\hat
v_{\pm}$ are not constant equal to one. But since we have chosen $v_{\pm}(z)$
to be monic polynomials of the same degree, $\hat
v_{\pm}$  converge to one for $\vert
w\vert\rightarrow\infty$. Then $\hat a\rightarrow 0$ and $\hat
\beta\rightarrow \pi$ brings our equations for $\vert w\vert\rightarrow\infty$
in the same form as in the $AdS_3$ case. Thus, to get a null $2n$-gon, we
expect as the appropriate boundary conditions at $\vert
z\vert\rightarrow\infty$ for the two coupled differential
equations (\ref{4gauss}) and (\ref{4ricci})
\beq
\alpha (z,\bar z)~=~(n-2)\log\vert z\vert ~+~o(1)~,~~~~~\beta (z,\bar
z)~\rightarrow \pi ~.\label{bc}
\eeq.
\section{Spacelike minimal surfaces in $AdS_5$}
Now the gauge group related to the normal space is non-Abelian
($SO(1,2\vert\mathbb{R})$), 
and the torsion invariant $T$ can have both signs. 

To identify gauge
invariant and holomorphic objects, we adapt the procedure used for $AdS_4$.
$A\in so(1,2\vert\mathbb{C})$ can be parameterized by some $M\in
SO(1,2\vert\mathbb{C})$ via \footnote{We skip the conjugated equations. For
 a mathematical comment on this parameterization see the second part of the
  appendix.}  
\beq A~=~\partial M~M^{-1}~.\label{5par}
\eeq 
Then 
\beq
v(z)~:=~\overline{ M^{-1}(z,\bar z)}~u(z,\bar z)\label{5vu}
\eeq
is holomorphic, due to (\ref{codazzi}). Gauge transformations
$u\mapsto \Omega u,~~M\mapsto \Omega M$ leave $v$ invariant, since
$\Omega\in SO(1,2\vert\mathbb{R})$ is real.  Again there remains
a gauge parameterization freedom
\beq
M(z,\bar z)~\mapsto ~M(z,\bar z)~\overline{Q(z)}~,\label{5repar}
\eeq
with holomorphic $Q(z)\in SO(1,3\vert\mathbb{C})$. Under such a transformation
$v(z)$ behaves as 
\beq 
v(z)~\mapsto ~Q^{-1}(z)~v(z)~.\label{5vtrafo}
\eeq
We have now identified three holomorphic functions\ : the three components of
$v(z)$, i.e. $v_4(z),v_5(z)$ and $v_6(z)$. The position of their
zeros has again invariant meaning under conformal coordinate transformations.
However, due to the unavoidable matrix structure in (\ref{5vtrafo}), there
is no invariant meaning of the zeros of all three holomorphic functions with
respect to a change of the gauge parameterization. One could try to use
(\ref{5vtrafo}) to set one or even two of the components of $v(z)$ to zero.
Away from the zeros of the old $v_k$ this is possible locally. But we did not
find a suitable way to implement some reduction with globally
holomorphic $Q$. The separate parameterizations of the gauge field used 
in \cite{djw} for $T<0$ and $T>0$ are not suitable to globally cover a
situation where $T$ changes sign on the surface. We suspect that sign
changes of $T$, which are not possible in $AdS_4$, are necessary to
realize the most generic boundary null $N$-gon in $AdS_5$. 

For sure there is one related holomorphic function whose zeros have invariant 
meaning, namely $v_kv^k$. Since $L=e^{-4\alpha}\vert v_kv^k\vert ^2$, these
zeros correspond to points, where the universal inequality (\ref{bound}) for
$R$ and $T$ is saturated. 

Irrespective of the outcome of counting the independent holomorphic functions,
there must be additional parameters beyond the position of their zeros.
From the zeros we get in any case an even number of real parameters.
However, for the 2n-gons in $\mathbb{R}^{1,3}$ one has $6n-15$ parameters.  
A resolution of this mismatch could come from free real parameters in the
non-Abelian gauge field solution, similar to scale parameters in instanton
solutions.\\ 

We close this section by a reformulation of the basic equations
(\ref{gauss})-(\ref{ricci}) for $AdS_5$, using the mapping of the
gauge group $SO(1,2)$ to $SL(2,\mathbb{R})$. This formulation so far
did not lead to further insights in the problem just discussed. We present
it here since it is interesting in its own right.

Choosing ($\sigma _k$ Pauli matrices) the following basis of $sl(2,\mathbb{R})$
\beq
\tau _0=-i~\sigma _2~,~~~\tau _1=\sigma _1~,~~~\tau _2=\sigma _3~,
\eeq
and (a shift $(4,5,6)\mapsto (0,1,2)$ understood)
\beq
U~=~u^k~\tau _k~,~~~~~~A~=~\frac{1}{4}~A^{kl}\tau _k\tau _l~,
\eeq
one gets
\beq
\partial\bar{\partial}\alpha ~-~\frac{e^{-\alpha}}{2}~\mbox{tr}\big (U\bar
U\big )~-~e^{\alpha}~=~0~,
\eeq
\beq
\bar{\partial}U~-~\big [\bar A,U\big ]~=~0~,~~~~~~~~~\partial \bar U~-~\big
[A,\bar U\big ]~=0~,
\eeq
\beq
e^{-\alpha}~\big [\bar U,U\big ]~=~\partial \bar A~-~\bar{\partial}A~+~\big [
\bar A,A\big ]~.
\eeq
In this form it is even more eye-catching, that the basic equations for the
reduced system in the sense of Pohlmeyer reduction, i.e. the set of Gau{\ss},
Codazzi and Ricci equations (\ref{gauss})-(\ref{ricci}), can be regarded as a
Hitchin system (A,U) \cite{hitchin}, coupled to the metric parameterized by
$\alpha$.  It
would be interesting 
to find a possibility to handle $\alpha$ as part of an enlarged gauge field,
so that the whole system is equivalent to a pure Hitchin system. Such a
construction has been realized for the $AdS_3$ case in \cite{am3}.

\section{Conclusions}
We discussed several geometrical issues related to the construction
of minimal surfaces with null polygonal boundaries at conformal infinity of
$AdS_m$. There are two independent scalar invariants, curvature $R$ and 
torsion $T$. They obey the universal inequality (\ref{bound}). Points where
this inequality is saturated, play a distinguished role for the surfaces
under consideration. 

For $AdS_3$ and $2n$-gons with $2n>4$, descriptive geometrical
arguments have been given for 
the existence of points with vanishing second fundamental form $u$. These were
based on the observation, that at points of a 
minimal surface with a crossing of three or more curves, which are geodesic in
$AdS_3$, $u$ has to vanish.

We introduced the notion of global conformal coordinates, which are
fixed up to translations and a multiplication with a constant. The existence
of such coordinates for the relevant minimal surfaces in $AdS_3$ is
guaranteed by the construction of Alday and Maldacena \cite{am3}.  
It is natural to assume their existence also for minimal surfaces in 
higher dimensional $AdS_m$ for null polygonal boundaries in the neighbourhood
of those of the $AdS_3$ type. 

In $AdS_4$ the Pohlmeyer reduced system for spacelike minimal surfaces 
can be reformulated such that it is described by two functions $\alpha (z,\bar
z),~\beta (z,\bar z)$, which obey the two coupled differential equations
(\ref{4gauss}),(\ref{4ricci}). In these equations appear two holomorphic
functions as input. Guided by the match of the number of parameters and the
similarities to the $AdS_3$ case, we formulated boundary conditions for
$\alpha$ and $\beta$. We conjectured, that with a polynomial ansatz for the two
holomorphic input functions the solution for the reduced system
and its boundary condition, after solving the related linear problem, yields
a null polygonal boundary. The conformal invariant data of the null polygon
are expected to be in one to one correspondence to the relative positions of
the zeros of the two holomorphic polynomials. 

In $AdS_5$ the torsion invariant for spacelike minimal surfaces can have
both signs. Also in this case it is straightforward to identify holomorphic
input functions. However, up to now we did not succeed in a full
identification of their gauge invariant content.\\[10mm]
{\bf Note added:}\\[2mm]
In two recent papers \cite{Jevicki:2009bv,Alday:2009dv} 
only $p(z)=\bar u_a\bar u^a$ has been kept as a holomorphic
function, and as the partner for $\alpha$ in the Pohlmeyer reduced
equations for $AdS_4$, instead of our $\beta$, a modified function has been
introduced via $\bar u_4=i \sqrt{ p(z)}\cos \tilde{ \beta}/2$, $\bar
u_5=\sqrt{p(z)}\sin \tilde{ \beta}/2$. Then the difference
$\tilde{\beta}-\beta$ is proportional to $\log\left (\frac{v_-\bar v_+}{\bar
    v_-v_+}\right )$ \cite{JJ,Alday:2009dv}, and thus contains some
winding in the $z$-plane. Winding for $\tilde{\beta}$ has been observed in 
appendix B.2. of \cite{Alday:2009dv} just in a way as required by the log-term
in $\tilde{\beta}-\beta$. This gives further support for our assumption 
that $\beta$ is regular everywhere. \\[10mm]    
\noindent
{\bf Acknowledgement}\\[2mm]
I thank George Jorjadze for many intense discussions and for reading the
manuscript. In addition, useful discussions with
Nadav Drukker, Chrysostomos Kalousios, Chri\-stoph
Meyer, Jan Plefka, Sebastian Wuttke and Donovan Young are acknowledged.
This work has been supported in part by Deutsche Forschungsgemeinschaft via
SFB 647.\\[20mm]
\noindent
{\Large\bf Appendix}\\[5mm]
We ask for curves on a minimal surface in $AdS_3$, which are geodesics in the
sense of the embedding space. At points with nonvanishing second
fundamental form at most two such curves can cross. If there is a point
with three or more such curves crossing, then at this point the second
fundamental form has to be zero. The analog statement for minimal surfaces
in $\mathbb{R}^3$ is obvious. To be on the safe side, we sketch here some
formulae proving it for our situation. 

Let $Y(t)$, $t$ affine parameter, describe a curve on the surface by its
coordinates in the embedding space of $AdS_3$, i.e. $\mathbb{R}^{2,2}$. Then
the curve is geodesic with respect to the surface iff
\beq
\ddot Y~=~\rho Y~+~\omega B_4
\eeq
and it is geodesic in $AdS_3$, iff in addition $\omega =0$.
Now we use $\frac{d}{d t}=\dot z\partial +\dot{\bar z}\bar{\partial}$
and (\ref{master}) to express $\ddot Y$ as a linear combination of
$Y,\partial Y, \bar{\partial}Y,B_4$. For a  geodesic  in the sense of the surface, the coefficients in front of
$\partial Y$ and $\bar{\partial}Y$
have to vanish, i.e.
\beq
\ddot z~+~\dot z^2~\partial\alpha~=~0~.
\eeq
To be in addition also geodesic in the sense of $AdS_3$, the coefficient
in front of $B_4$ has to vanish, i.e.
\beq
\dot z^2~u~+~\dot{\bar z}^2~\bar u~=~0~.
\eeq
Expressed in terms of the real coordinates $\sigma$ and $\tau$ and using
$u=a+bi$, the last equation becomes
\beq
(\dot{\sigma},\dot{\tau})~\left (
\begin{array}{ll}
~~a&-b\\
-b&-a
\end{array}\right )~\left (
\begin{array}{l}
\dot{\sigma}\\
\dot{\tau}
\end{array}\right )~=~0~.\label{geod}
\eeq
The eigenvalues of the matrix are $\pm\sqrt{a^2+b^2}$. Hence as long as
$u\neq 0$, there are just two (orthogonal) directions in the $z$-plane
which satisfy (\ref{geod}).\\

Finally, a comment on the parameterization (\ref{5par}) is in order. It is
obvious that 
it generates elements of $so(1,2\vert\mathbb{C})$. Since $\partial $ is a
complex derivative, it is less obvious that
all elements of this Lie algebra can be represented in this manner.  
The differential equation (\ref{5par}) is equivalent to the integral equation
\beq
M~=~\frac{1}{\partial\bar{\partial}}~\bar{\partial}(AM)~+~\mathbb{I}~.
\label{int}
\eeq 
A solution of this equation as an infinite series can be generated by
successive approximation, starting with zeroth approximation
$M^{(0)}=\mathbb{I}$. Up to now we did not find a proof, that the resulting
matrix is indeed out of $SO(1,2\vert\mathbb{C})$.

\end{document}